\documentclass[pra,twocolumn,aps,showpacs,amsmath,amssymb,floatfix]{revtex4-1}

\usepackage{color}
\usepackage{bm}
\usepackage{graphicx}
\usepackage{braket}

\usepackage[plainpages=false,pdfpagelabels,colorlinks=true,linkcolor=red,urlcolor=blue,citecolor=blue,pdftitle={Title},pdfauthor={},pdfdisplaydoctitle=true,pdfduplex=DuplexFlipLongEdge]{hyperref}

\newcommand\ba{\begin{eqnarray}}
\newcommand\ea{\end{eqnarray}}
\newcommand\be{\begin{equation}}
\newcommand\ee{\end{equation}}

\begin{document}

\title{Geometric quenches in  quasi-disordered lattice system }
\author{Ravi Kumar$^1$ and Ranjan Modak$^2$} 
\affiliation{$^1$ Department of Physics, Indian Institute of Technology (Banaras Hindu University), Varanasi 221005, India}
\affiliation{$^2$ Department of Physics, Indian Institute of Technology Tirupati, Tirupati 517506, India}

\begin{abstract}
While global quantum quench has been extensively used in the literature to understand the localization-delocalization transition for the one-dimensional quantum spin chain, the effect of geometric quench (which corresponds to a sudden change of the geometry of the chain)  in the context of such transitions is yet to be well understood. In this work, we investigate the effect of geometric quench in the Aubry-Andre model, which supports localization-delocalization transition even in one dimension. We study the spreading of the entanglement and the site-occupation with time and find many interesting features that can be used to characterize localization-delocalization transition. We observe that geometric quench causes a power-law type growth of the entanglement entropy in the delocalized phase in contrast to the linear growth which is found in the global quench studies. Remarkably, we also find that the saturation values in the Many-body localized (MBL) phase obey Area law in contrast to the usual volume law which is a signature feature of MBL phase in the context of global quench. 

\end{abstract}
\maketitle

\section{Introduction}
In one dimension any arbitrary weak amount of disorder is sufficient to localize all eigenstates of a non-interacting system. This phenomenon is famously known as Anderson localization ~\cite{anderson.58,tvr.79}. The question of how this picture is modified by interactions remained unclear for a very long time. However, relatively recently, Basko, Aleiner, and Altshuler have argued that an interacting many-body system
can undergo a so-called many-body localization (MBL)
transition in the presence of quenched disorder~\cite{basko.2006}. There have been a plethora of work in this direction in last one decade to understand the nature of this phase transition both theoretically ~\cite{mbl1,mbl2,mbl3,mbl4,mbl5,mbl6,mbl6,mbl7,mbl8,mbl8,mbl9,mbl10,mbl14}, and experimentally as well ~\cite{mbl11,mbl12,mbl13}.
The MBL transition is rather unique in contrast to more conventional quantum phase transitions. This is not a transition in the ground state, instead, the MBL transition involves the localization of highly excited states of a many-body system, with finite energy density. Also this
MBL phase is of fundamental interest in the context
of statistical mechanics. Local subsystems of a generic interacting 
many-body system are expected to equilibrate with their surroundings, and that has led to the so-called eigenstate thermalization
hypothesis (ETH), which states that individual eigenstates of the interacting system encode
thermal distributions of local quantities ~\cite{eth1,eth2,eth3}. However,
the many-body localized phase is an exception, in which the individual eigenstates fail to obey ETH, and the notion of ergodicity breaks down ~\cite{meth1,meth2}.

Most of the  out-of-equilibrium studies of localization-delocalization transition involve
the so-called quantum quench, in which a system is initially
prepared in the ground state of a many-body quantum Hamiltonian, and a non-trivial unitary dynamics is then induced by changing instantaneously (i.e., quenching) one (or many)
control parameters. Depending on whether this change happens locally or in the whole system, the quench falls into the class of local or global quenches, respectively. One uses universal features of  different diagnostics e.g. entanglement entropy, out of time correlators (OTOC) to distinguish between different phases of the systems ~\cite{ent1,ent2,ent3,ent4,ent5,ent6}.

In our work, we focus on a situation that is intermediate between a local and a global quench. We consider the real-time dynamics following an instantaneous change of
the geometry or the size of the system, the so-called geometric quench~\cite{gm1,gm2,gm3,gm4}. More specifically,  we prepare an initial state which is say the ground state of a lattice Hamiltonian of length $L_A$, then study unitary dynamics under the same lattice Hamiltonian of length $L$, where $L> L_A$. The question we are addressing here is whether one can use this geometric quench as a probe to detect localization-delocalization transitions. Thanks to extraordinary advancements of ultra-cold gas experiments, this kind of sudden expansion of lattice size or traps have been realized in recent days ~\cite{sudden_expt1,sudden_expt2}.

 Given that non-interacting one-dimensional  system of fermions  in presence of a true disorder does not show any localization-delocalization transition, Aubry-Andre (AA) Hamiltonian ~\cite{Aubry80} is one of the best suited Hamiltonian in order to investigate the effect of geometric quench in the localization-delocalization transition.
 Instead of pure randomness this model has the incommensurate on-site potential which drives a system in the localized phase and the localization-delocalization transition occurs for a finite incommensurate potential amplitude 
 in contrast to the usual Anderson localization in one-dimension, which requires only an infinitesimal disorder strength to localize all states. One can also explore the MBL transition introducing the interaction in this model~\cite{iyer.2013}.
 
 In our study, we investigate the effect of geometric quench in the Aubry-Andre model. We have used the spreading of the entanglement and the site-occupation with time, as two tools to characterize localization-delocalization transition. 
 Our main two striking results are the following.
 1) We have observed that geometric quench causes a power-law type growth of the entanglement entropy in the delocalized phase in contrast to the linear growth which is found in the global quench studies. 
 2) The saturation values in the Many-body localized (MBL) phase obey Area law in contrast to the usual volume law which is a signature feature of MBL phase in the context of global quench~\cite{moore.2014}.

The paper is organized as follows. In Sec. II, we introduce the  model and protocols. Next we discuss the characteristics of site-occupation profile in Sec. III. In Sec. IV
we  investigate entanglement dynamics followed by Geometric quench. Finally, in Sec. V we summarize our results.

\section{Model and Protocols}
We study a system of fermions  in an one-dimensional lattice of size $L$, which is described by  the following  Hamiltonian:
\begin{eqnarray}
\hat{H}&=&-\sum_{i=1}^{L-1}(\hat{c}^{\dag}_i\hat{c}_{i+1}+\text{H.c.})+2h\sum_{i=1}^{L} \cos(2\pi\alpha i+\phi)\hat{n}_i  \nonumber \\
&+&V\sum_i\hat{n}_i\hat{n}_{i+1},
\label{nonint_model}
\end{eqnarray}
where $\hat{c}^{\dag}_i$  ( $\hat{c}_i$) is the fermionic creation (annihilation) operator at site $i$, $\hat{n}_i =\hat{c}^{\dag}_i\hat{c}_{i}$ is the number operator, and  $\alpha$ is an irrational number. Without loss of any generality, we choose $\alpha=\frac{\sqrt{5}-1}{2}$ and $\phi$ is a random number chosen between $[0,2\pi]$. We do averaging over $\phi$ for all the calculations presented in this work to obtain  better statistics. 
In the absence of interaction i.e. $V=0$, the Hamiltonian $\hat{H}$ is known as Aubry-Andr{\'e} (AA) model. It  supports a  delocalization-localization transition as one tunes $h$. In the thermodynamic limit, $h=1$ corresponds to the transition point \cite{Aubry80}.

Given that we wanted to investigate the effect of Geometric quench in these systems, we do the following quench protocols. 
First, we prepare the initial state as a ground state of the Hamiltonian $\hat{H}$ in a one-dimensional lattice of size $L_A < L$, and keep the $(L-L_A)$ sites completely empty. 
For all our calculations We fixed the total number of fermions as $N=L_A/2$ and $L_A=L/2$. Then we let the state evolved under the unitary evolution of the Hamiltonian $\hat{H}$ which is supported in an one dimensional lattice of size $L$. This quench protocol is different than the local-quench ~\cite{cc2007,ep2007,ep2008,sd2011}, where the initial state is obtained by "gluing" together two identical copies of the ground state i.e. $|\psi (t=0)\rangle=|GS\rangle _{L_A} \otimes |GS \rangle_ {(L-L_A)}$. On the other hand, in our case the initial state is chosen to be, 
$|\psi (t=0)\rangle=|GS\rangle _{L_A} \otimes |0 \rangle_ {(L-L_A)}$. For $V=0$, all the calculations are done using on-body density matrix approach~\cite{rigol.2005},  while for interacting case we use finite time density matrix renormalization group (tDMRG) technique ~\cite{dmrg1,dmrg2,dmrg3} to obtain all the results. 
Some of the data displayed in the main text are obtained using the tDMRG algorithm, as implemented in the ITensor Julia library~\cite{dmrg4}.


\begin{figure}
    \centering
    \includegraphics[width=0.48\textwidth]{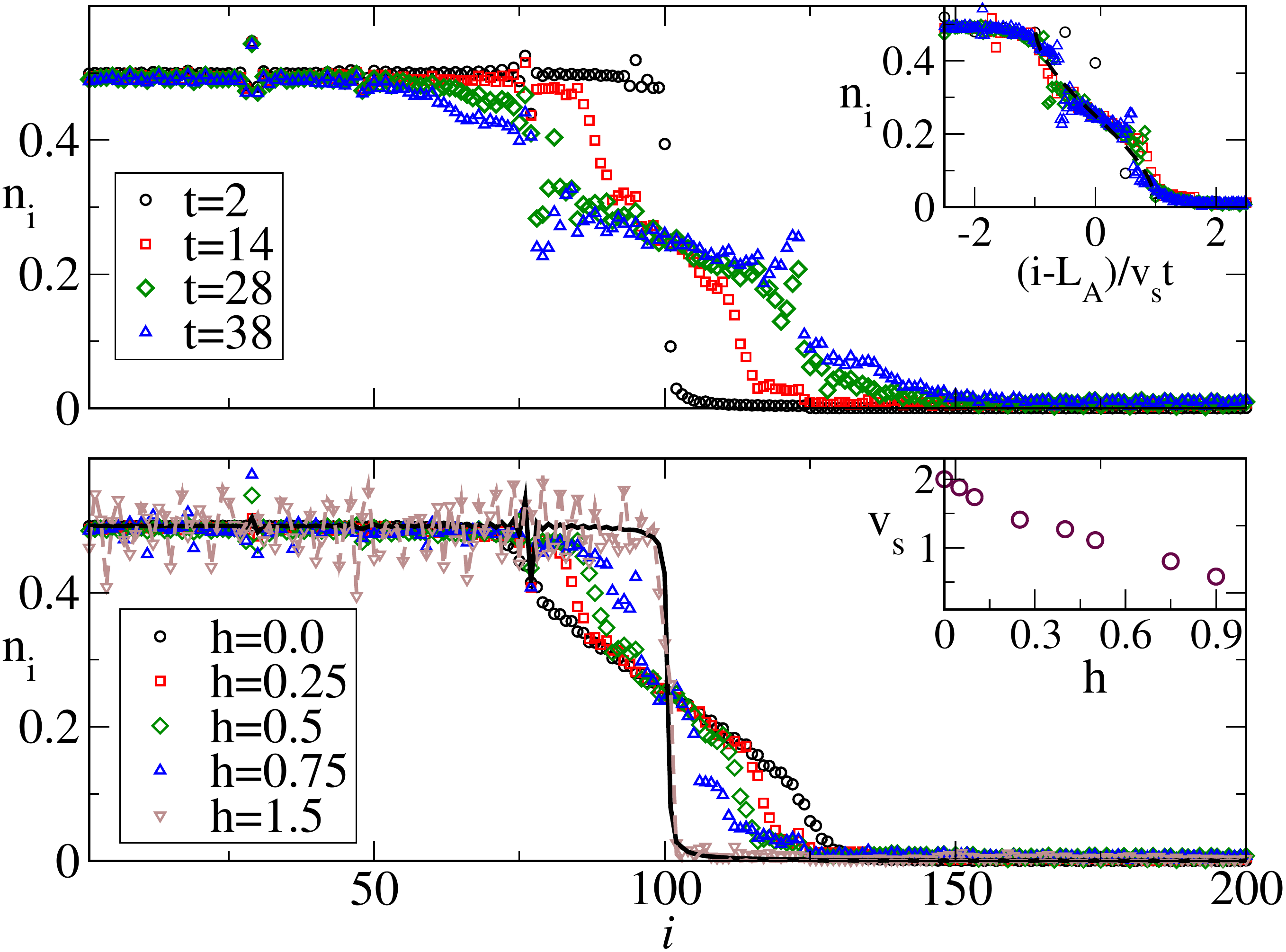}
    \caption{(Upper panel) Shows the variation of the site occupation for different time 
    for $h=0.5$, and $L=2L_A=200$. Inset shows the data collapse for the data for different times  as we rescale the x axes with $(i-L_A)/v_st$ and the dashed line represents 
    $n_i=1/4-(1/2\pi)\sin ^{-1}[(i-L_A)/v_st]$, where $v_s$ is the fitting parameter.
    (Lower panel) Shows the variation of $n_i$ for different values of $h$ for fixed time $t=14$. Solid and dashed lines corresponds to the site occupation for $t=0$ for $h=0.25$ and $h=1.5$ respectively. 
    Inset shows the variation the fitting parameter $v_s$ with $h$.}
    \label{fig1ab}
\end{figure}

\begin{figure}
    \centering
    \includegraphics[width=0.48\textwidth]{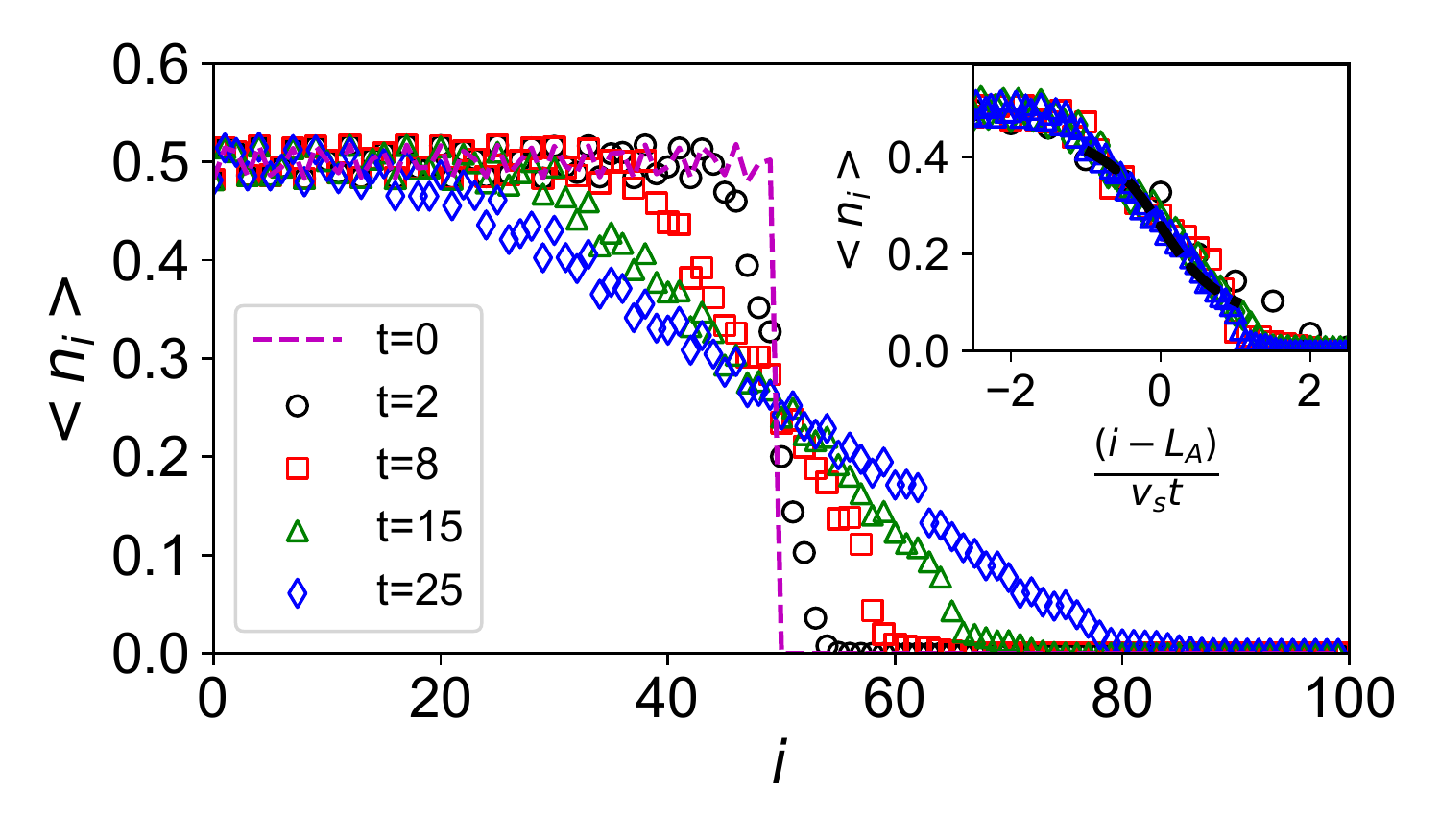}
    \includegraphics[width=0.48\textwidth]{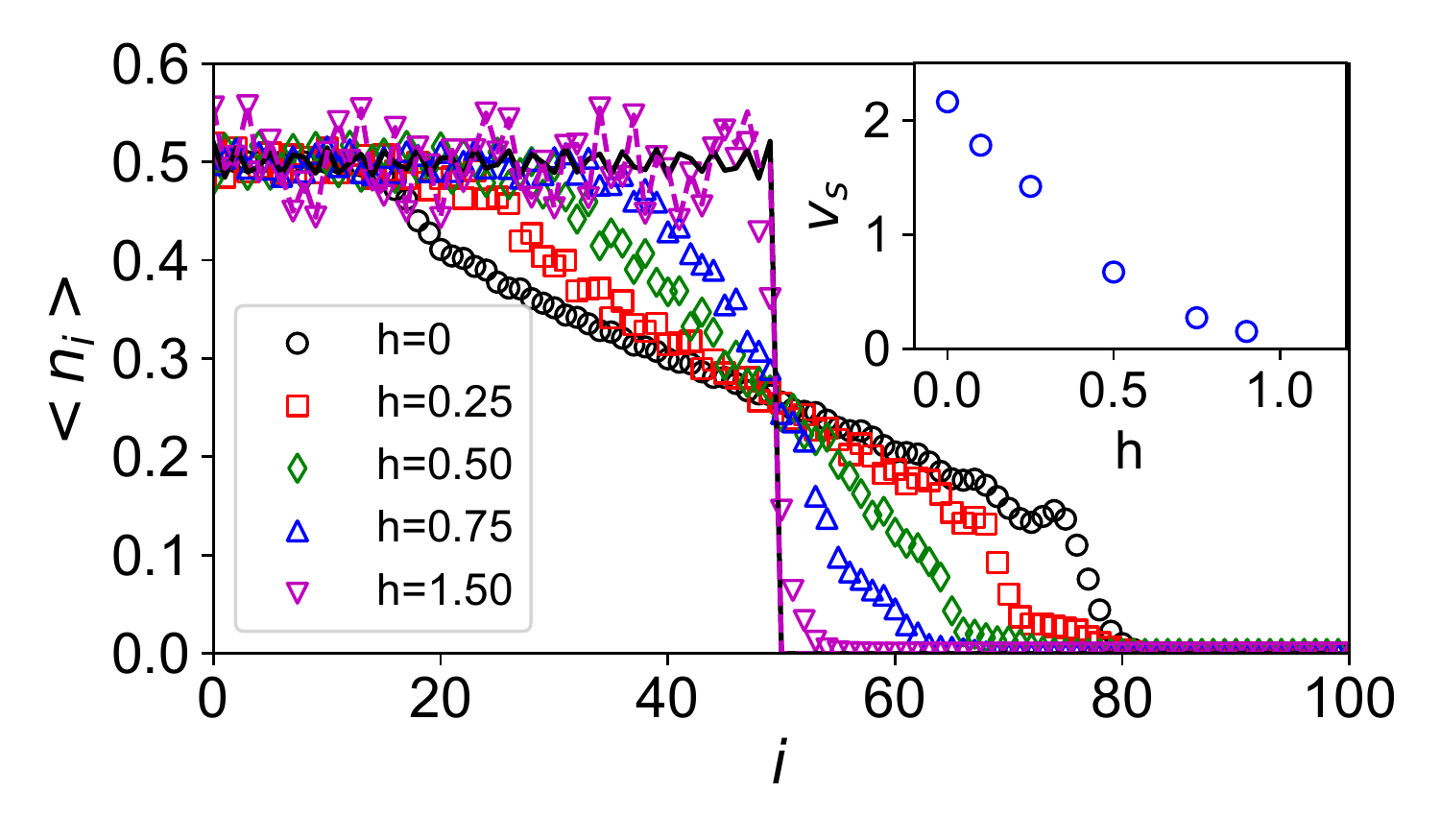}
    \caption{(Upper panel) Shows the variation of the site occupation for different time for $h=0.5$,$V=0.5$ and $L=2L_A=100$. Inset shows the data collapse for the data for different times  as we rescale the x axes with $(i-L_A)/v_st$ and the black dashed line represents 
    $n_i=1/4-(1/2\pi)\sin ^{-1}[(i-L_A)/v_st]$, where $v_s$ is the fitting parameter.
    (Lower panel) Shows the variation of $n_i$ for different values of $h$ for fixed time $t=15$ for $V=0.5$. Solid and dashed lines corresponds to the site occupation for $t=0$ for $h=0.0$ and $h=1.5$ respectively. 
    Inset shows the variation the fitting parameter $v_s$ with $h$.}
    \label{fig2ab}
\end{figure}


\begin{figure}
    \centering
    \includegraphics[width=0.48\textwidth]{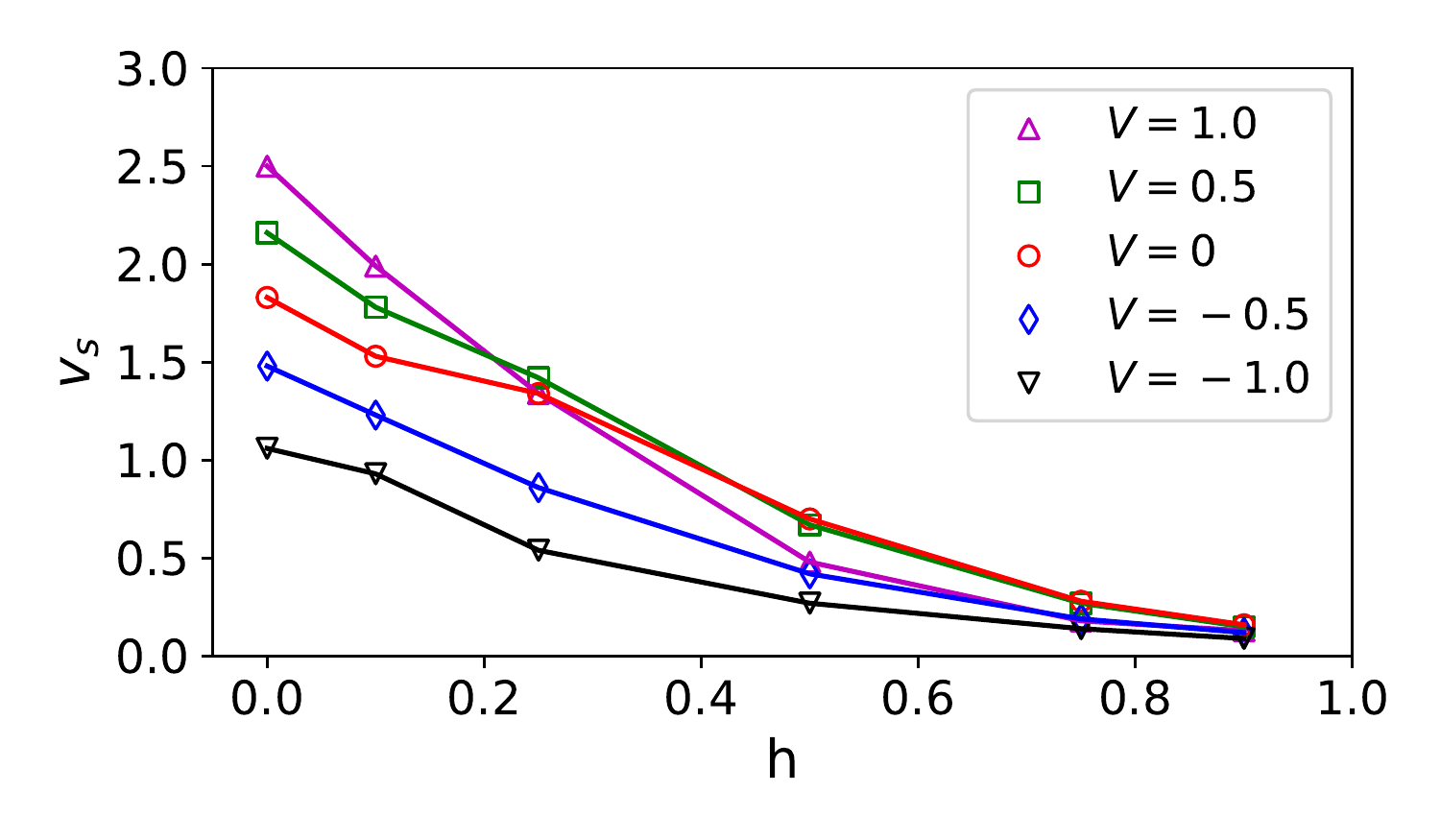}
    \caption{Shows the variation of the $v_s$ with  the disorder strength for $V=1, 0.5, 0, -0.5, -1.$ }
    \label{fig4}
\end{figure}

\section{Site occupation}
In this section, we discuss the real-time dynamics of the site occupation $n_i=\langle \hat{n}_i \rangle$ subject to the tuning of disorder strength ($h$) and the interaction ($V$) after the geometric quench. First, we prepare the initial state to be a ground state of a Hamiltonian ~\eqref{nonint_model} on a lattice of size $L/2$. Then we attach an empty lattice of size $L/2$ with it. Hence, the site occupation has a domain wall profile
in the beginning. Then the site occupation wavefront propagates with a velocity $v_s (V,h) $ (function of disorder strength and the interaction) for $t > 0$.

We first focus on the non-interacting case i.e. $V=0$. Figure.~\ref{fig1ab} (upper panel), shows the evolution of site occupation $n_i=\langle \hat{n}_i (t) \rangle$ at different time steps for $h=0.5$. Given that for $h=0.5$ the Hamiltonian ~\eqref{nonint_model} remains in the delocalized phase, one expects that the wave-front would propagate towards the boundary of the lattice, that is precisely what is observed in Fig.~\ref{fig1ab}. The next question one should ask is how to evaluate the wavefront propagation velocity $v_s$?
For that we use the ansatz i.e. $\langle n_i(t)\rangle=1/4-(1/2\pi)\sin ^{-1}[(i-L_A)/v_st]$, which
can be obtained analytically using a semiclassical reasoning that was also applied in Refs~\cite{eq1,eq2} for $h=0$.
We also get a remarkable data collapse for $n_i=\langle \hat{n}_i \rangle$ versus re scaled variable $(i-L_A)/v_st$ as shown in the inset of Figure.~\ref{fig1ab}. 

Next, we discuss the effect of disorder strength on the propagation of site occupation wavefront. As we increase the disorder strength ($h$), the velocity of the wavefront starts decreasing, and finally, the wavefront almost gets frozen as we cross the transition point i.e. $h=1$ as shown in Fig.~\ref{fig1ab} (lower panel). Inset shows the variation of $v_s$ with $h$.
Now we focus on the effect of interactions. Once we switch on the interaction, we observe melting of domain wall in site occupation profile similar to the one observed earlier for the non-interacting case. Figure.~\ref{fig2ab}  (upper panel)  describe the propagation of site occupation wavefront at different time steps for $V=0.5$. Once again we use the same ansatz i.e. $\langle n_i(t)\rangle=1/4-(1/2\pi)\sin ^{-1}[(i-L_A)/v_st]$ to extract the $v_s$. Figure.~\ref{fig2ab}  (lower panel) shows the change in behaviour of site occupation profile as a function of $h$ for a given time i.e. $t=15$. As expected, with the increase of $h$, the velocity of the propagating wavefront starts decreasing as shown in the inset of Fig.~\ref{fig2ab} (lower panel). In order to make the comparison even more clear, we plot the variation of $v_s$ with $h$ in Fig.~\ref{fig4} for different values of interaction
strength. We see a general feature with $v_s$ that dies down as we increase the disorder strength $h$ even for the interacting case. This is due to the effect of the many-body localization effect, as we approach the ergodic-MBL transition point the site occupation profile hardly changes with time. 
\begin{figure}
\vspace{0.4 in}
    \centering
    \includegraphics[width=0.45\textwidth, height=0.28\textwidth]{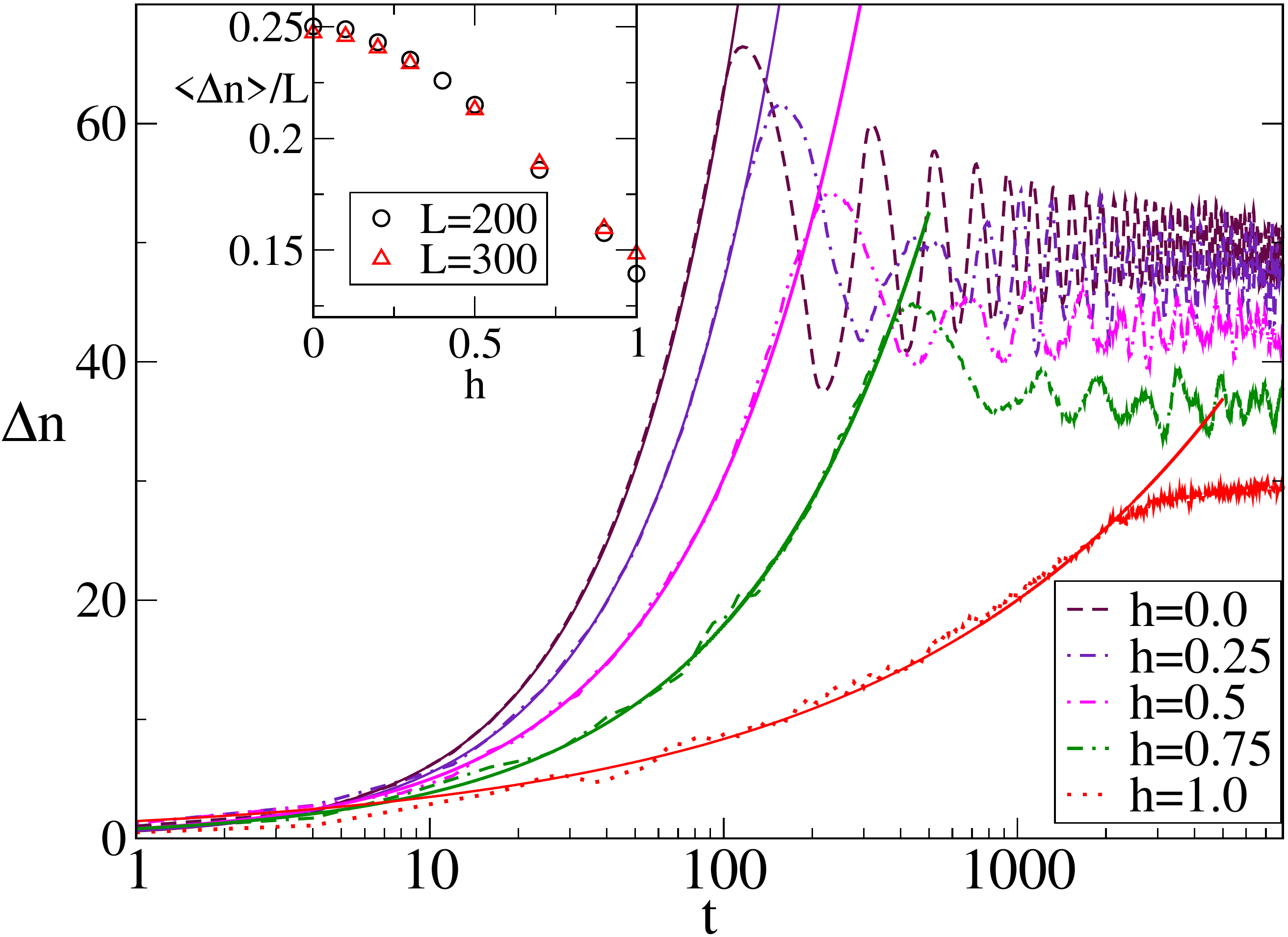}
    \caption{(Main panel) Shows the variation of $\delta n=\sum _i|n_i(t)-n_i(t=0)|$ with  time for different values of $h$ and $V=0$, and $L=2L_A=200$. 
    Solid lines is $\Delta n\sim t^{\gamma}$ with $\gamma=1, 0.93, 0.78, 0.67, 0.38$ for $h=0, 0.25, 0.5, 0.75, 1$ respectively. 
    Inset shows the variation for long time average of $\Delta n /L$ with $h$ for $L=200$ and $300$. }
    \label{fig2}
\end{figure}
In order to quantify the change in the site occupation profile with time in the delocalized phase, we use another quantifier i.e. $\Delta n (t)=\sum _i |n_i(t)-n_i(t=0)|$. In Fig.~\ref{fig2}, we show the variation of $\Delta n(t)$ 
with $t$ for non-interacting case for different values of $h$. We find an initial power-law growth i.e.  $\Delta n\sim t^{\gamma} $
and then it saturates (apart from some small oscillations). Interestingly this exponent $\gamma\simeq 1$ for $h=0$, but it decreases as one increases the value of $h$. In the inset we also show how the long time average of $\Delta n$, i.e. $\langle \Delta n \rangle=\frac{1}{T_2-T_1}\int_{T_1}^{T_2} {\Delta n dt}$ (note that we choose $T_1$ and $T_2$ to be large so that $\Delta n$ at that time window remains in the saturation regime), for different values of $h$ and $L$ and we find that $\langle \Delta n\rangle$ obeys a volume-law. 
\begin{figure}
    \centering
    \includegraphics[width=0.48\textwidth]{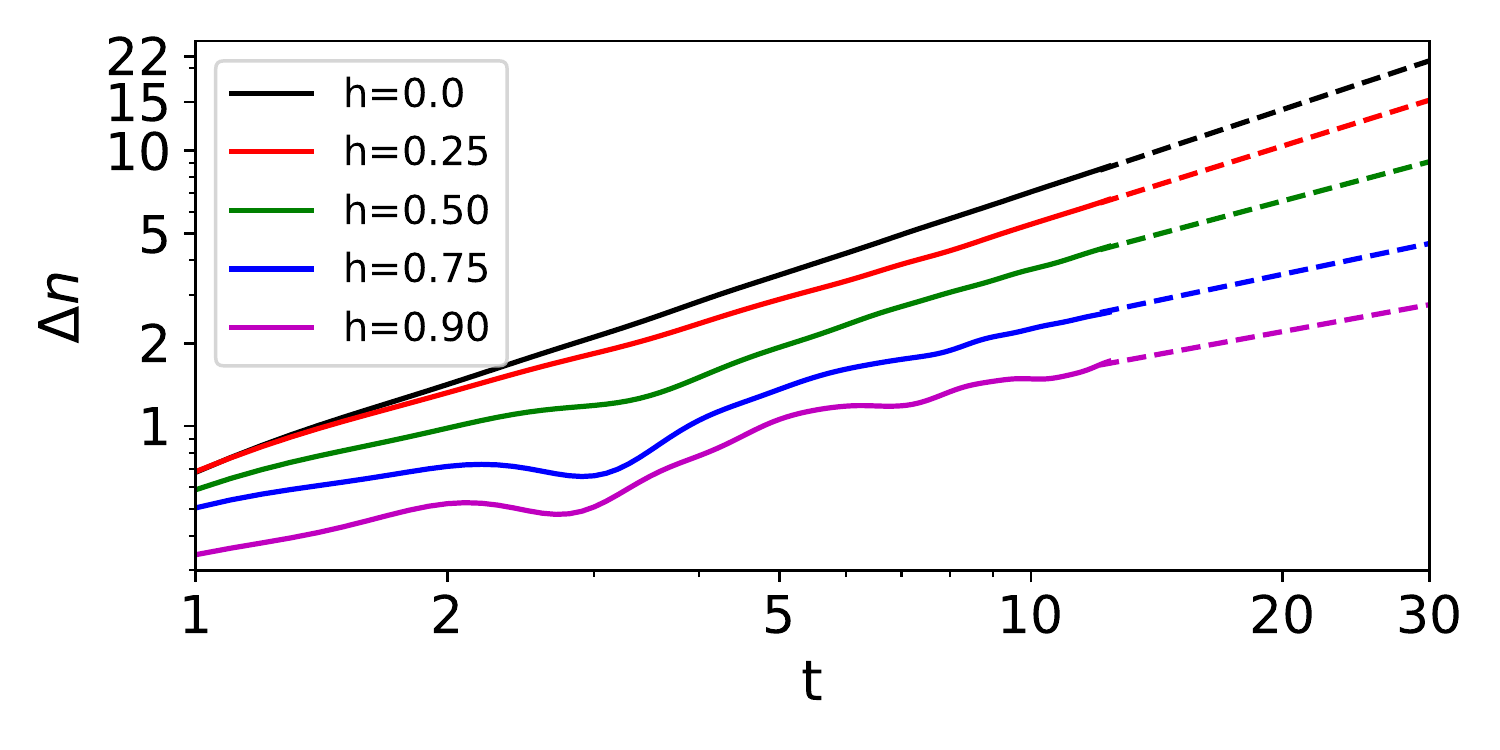}
    \includegraphics[width=0.48\textwidth]{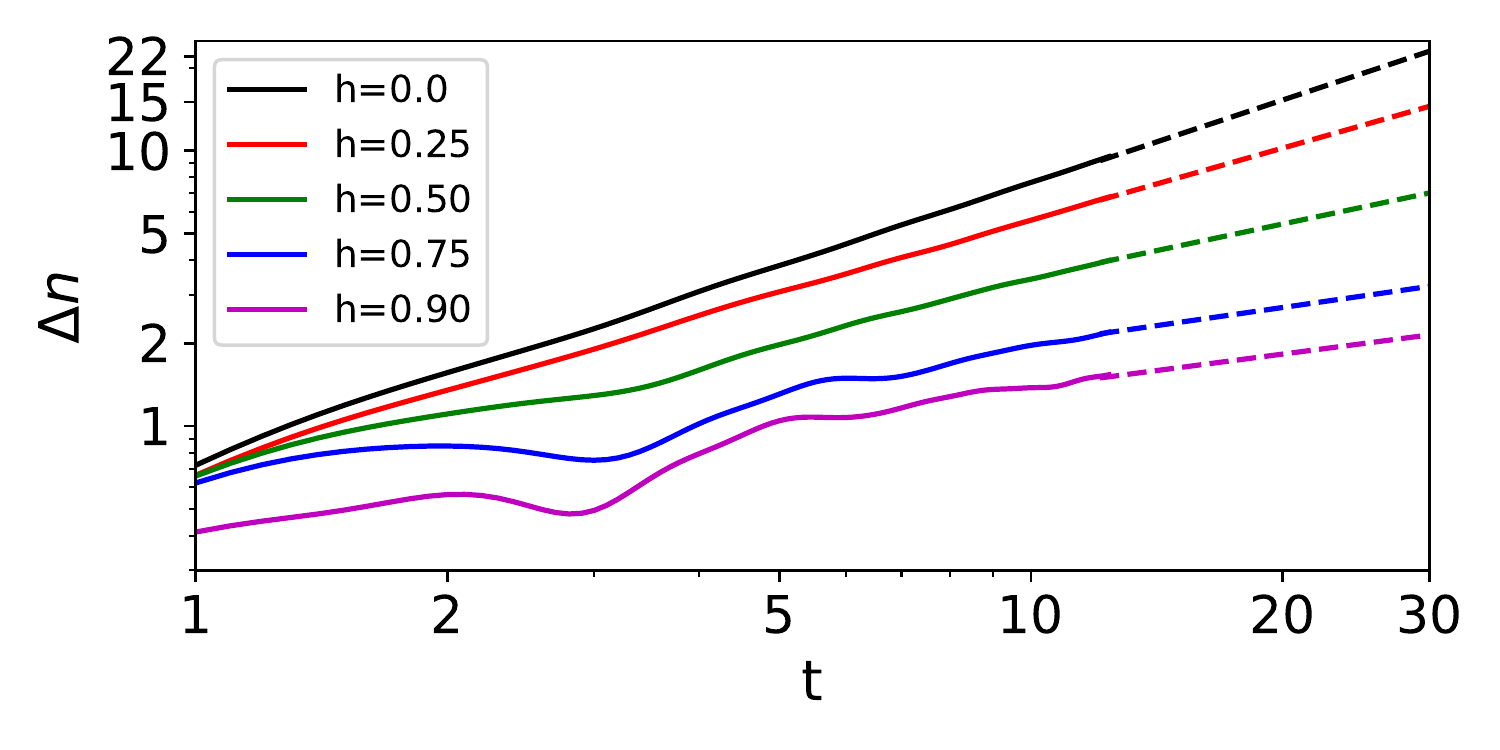}
    \caption{Shows the variation of $\delta n=\sum _i|n_i(t)-n_i(t=0)|$ with  time for different values of $h$ and the interaction $V$. Top figure shows the variation for $V=0.5$ where solid lines is $\Delta n\sim t^{\gamma}$ with $\gamma=1.00,0.94,0.80,0.63,0.55 $ for $h=0, 0.25, 0.5, 0.75, 0.90 $ respectively \\
    Bottom figure shows the variation for $V=1.0$ where solid lines is $\Delta n\sim t^{\gamma}$ with $\gamma=1.00,0.85,0.63,0.43,0.39$ for $h=0, 0.25, 0.5, 0.75, 0.90 $ respectively }
\label{fig5}
\end{figure}
Next, we investigate the effect of interactions. As Fig.~\ref{fig5} suggests, we find a similar behavior of $\Delta n(t)$ with $t$ even in presence of interaction.

\begin{figure}
    \centering
    \includegraphics[width=0.48\textwidth]{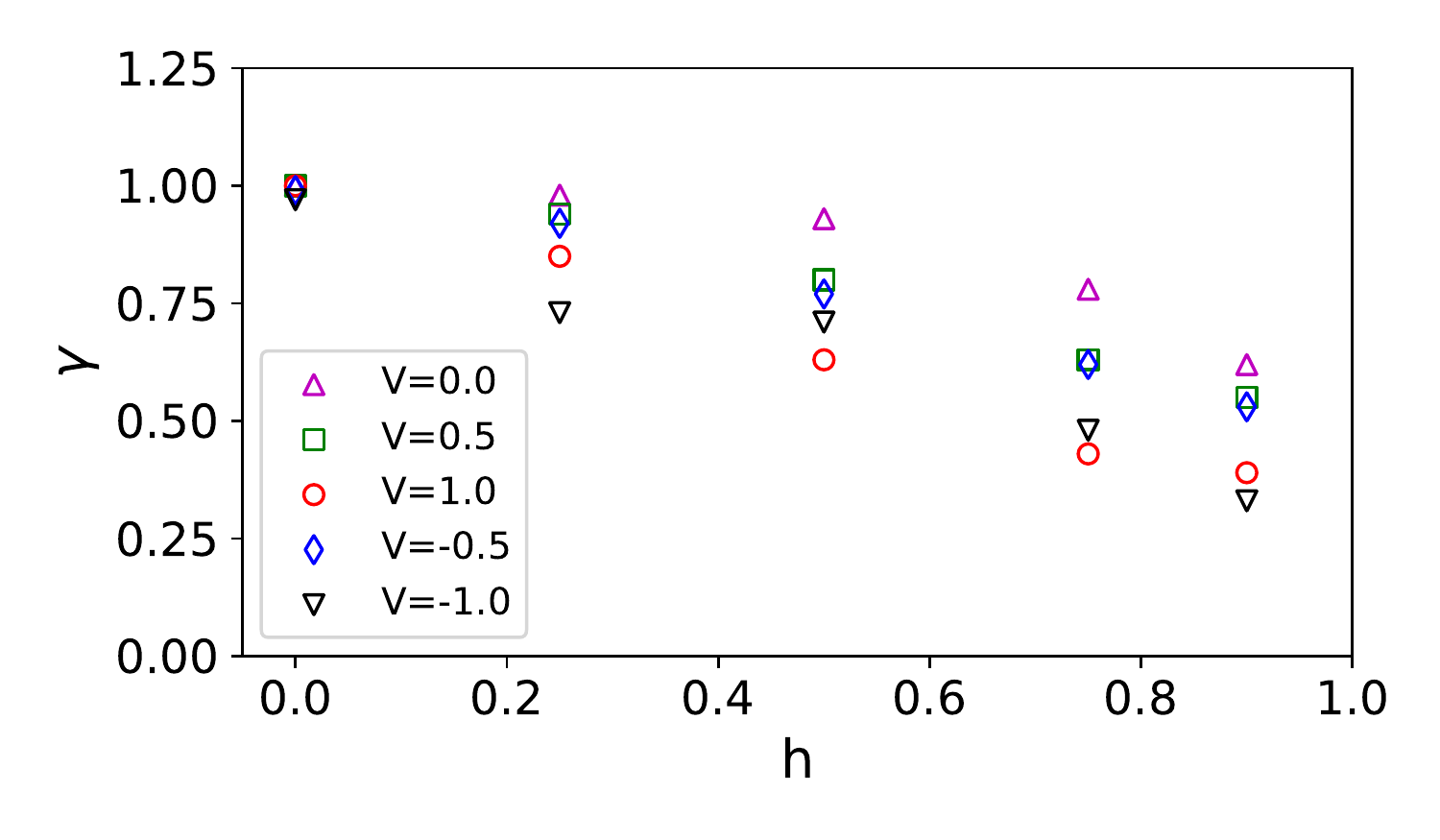}
    \caption{Scaling exponent $\gamma$ with the interaction and the disorder strength}
\label{fig6}
\end{figure}

However, the exponent $\gamma$
depends on the interaction and disorder strength which has been displayed in the
Fig.~\ref{fig6}. While for $h=0$, it seems that $\gamma$ does not depend on interaction strength significantly, but for non-zero $h$, there is a generic trend that the value of $\gamma$ (for a given $h$) decreases as the magnitude of interaction strength increases.   Due to the limitation of tDMRG simulation we were unable to reach a very long time, hence the saturation value of $\Delta n$ can't be analyzed as one could do it for the non-interacting case.  
\begin{figure}
    \centering
    \includegraphics[width=0.48\textwidth]{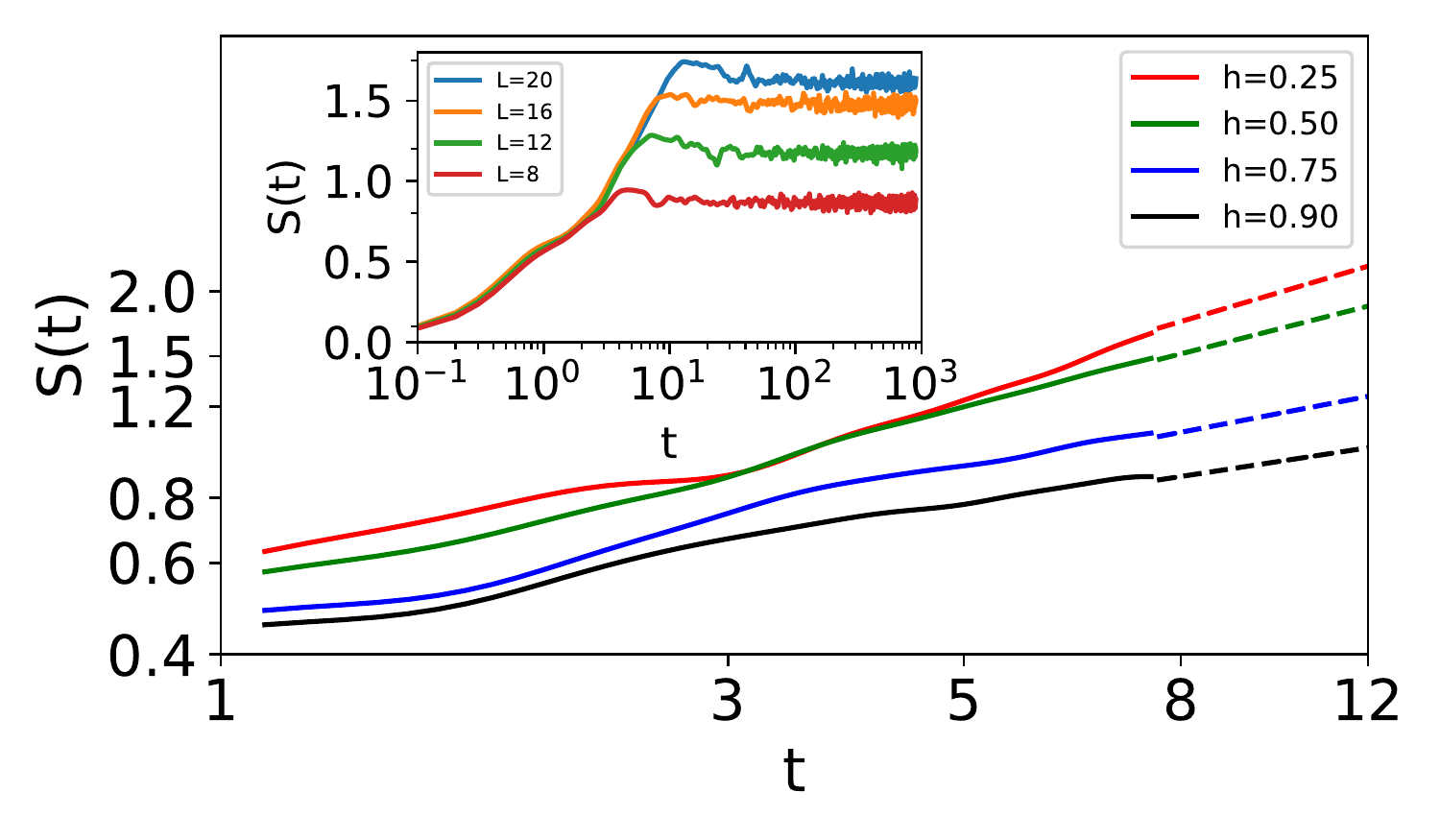}
    \includegraphics[width=0.48\textwidth]{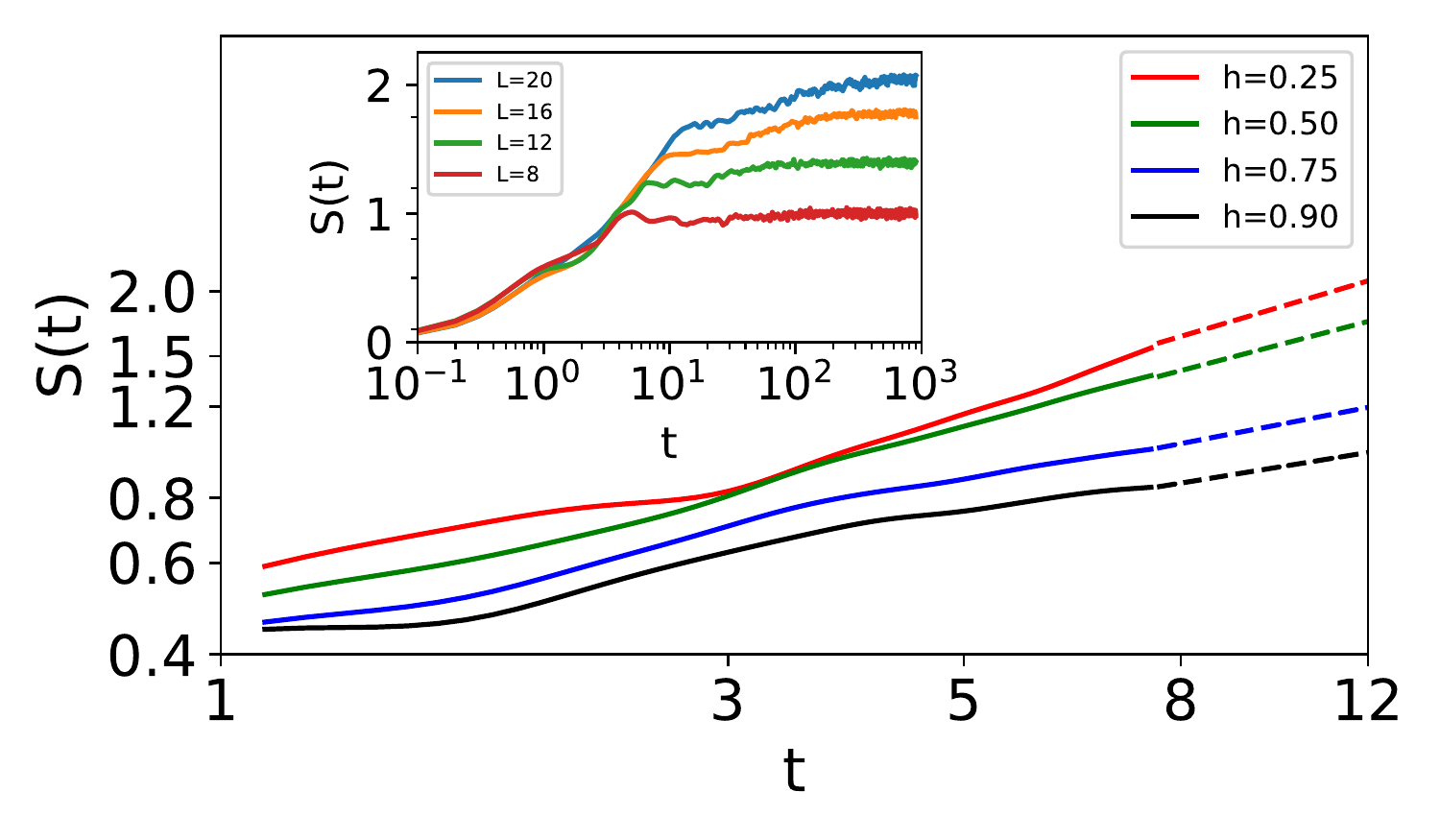}
   
    \caption{Evolution of entanglement entropy in the de-localised regime.
    Solid lines represents actual data and dotted lines is best fit line for $S(t) \sim t^{\eta}$ for different values of $h$ .\\ Top figure shows the variation for $V=0$ of system size $L=32$ where $\eta=0.60,0.52,0.39,0.31$ for $h= 0.25, 0.5, 0.75, 0.90 $ respectively. Inset shows the saturation of the entanglement entropy   for system sizes ,$L=8,12,16,20$ for $h=0.5$  \\
    Bottom figure shows the variation for $V=0.5$ of system size $L=32$ where $\eta=0.61,0.54,0.40,0.34$ for $h= 0.25, 0.5, 0.75,0.90 $ respectively. Inset shows the saturation of the entanglement entropy   for system sizes ,$L=8,12,16,20$ for $h=0.5$ }
\label{fig7}
\end{figure}
\begin{figure}
    \centering
    \includegraphics[width=0.48\textwidth]{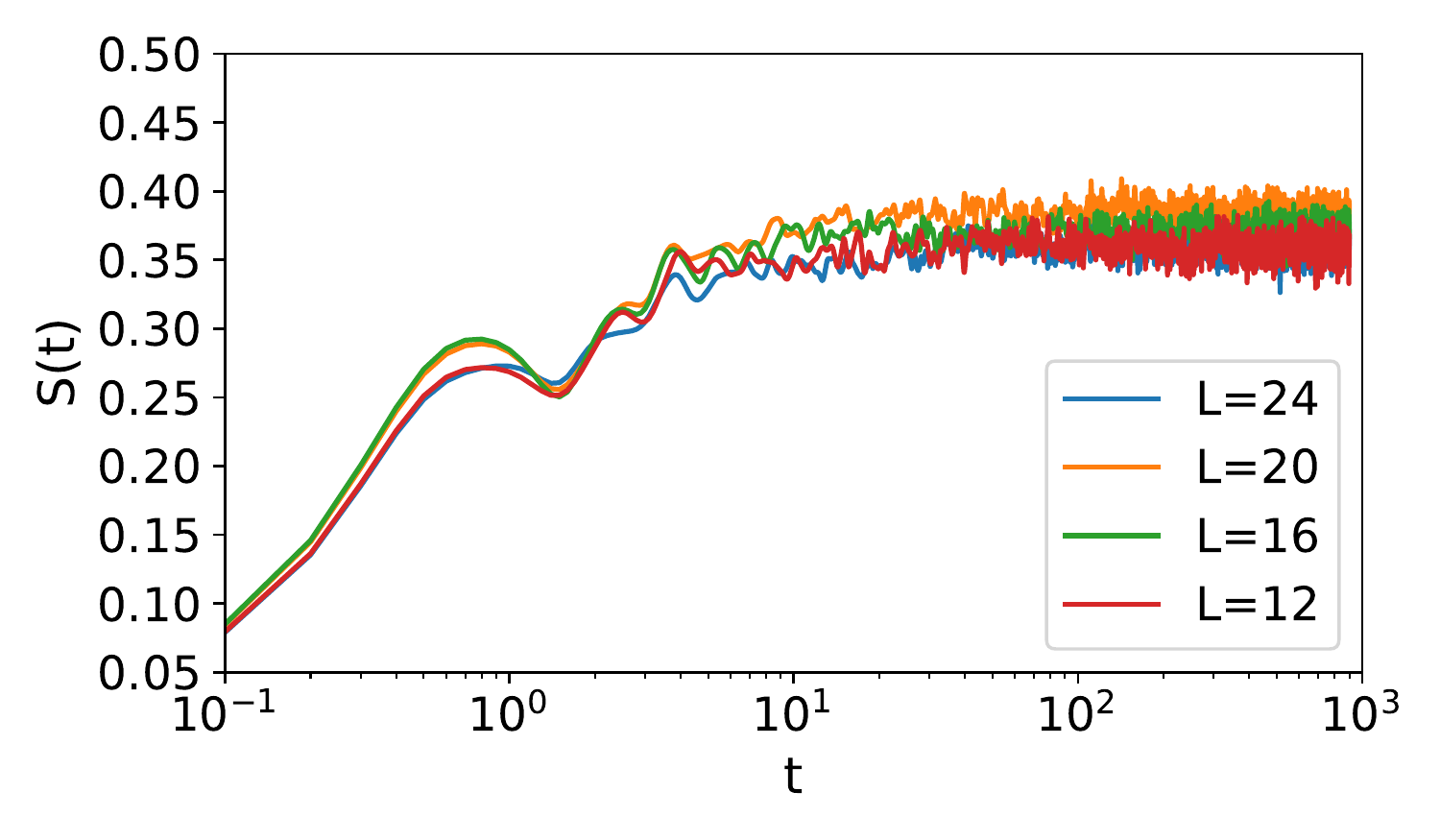}
    \includegraphics[width=0.48\textwidth]{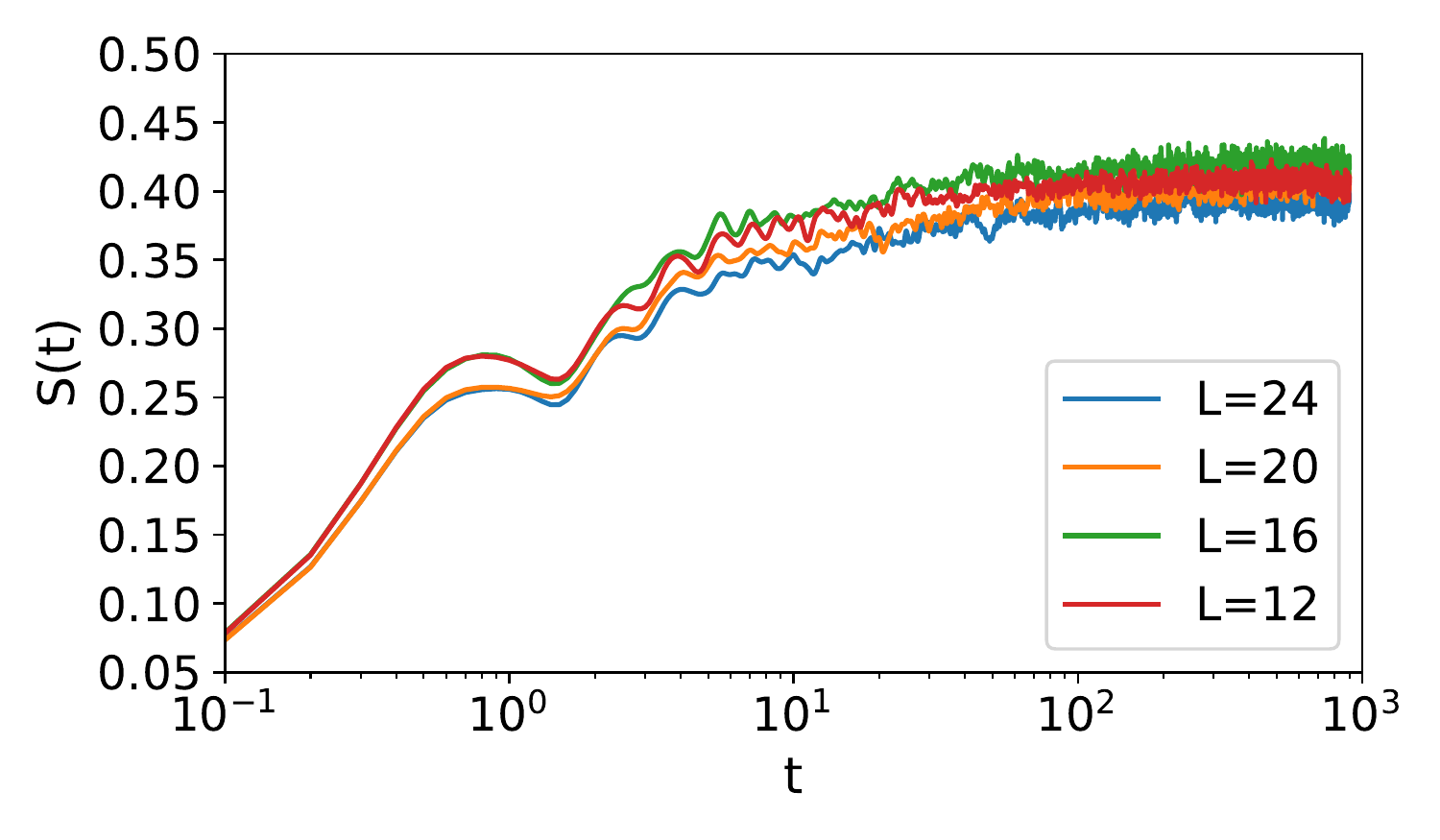}
       \caption{Evolution of entanglement entropy in the localised region ( h = 2.0 ) for $ L=  12,16,20,24$ for V = 0 (top) and V = 0.5(bottom). }
\label{fig8}
\end{figure}
\section{Entanglement dynamics}
Next, we investigate the entanglement dynamics. Even though there are many measures to characterize the entanglement, here we focus on probably one of the most popular measures of the entanglement i.e.  Von-Neuman entanglement entropy. Considering
a bi-partition of a system that is in a pure state  $|\psi\rangle$ into parts
A and B, a standard measure of their mutual entanglement is
the von Neumann entropy $S_A= -\text{Tr}\rho_A \log \rho_A$. 
Here $\rho_A$ is the reduced density matrix for A, obtained after
tracing part B from the full density matrix $\rho =|\psi\rangle \langle \psi|$.

We present the results for the time evolution of the entanglement entropy following the geometric quench in Fig.~\ref{fig7}. First, we focus on the delocalized phase i.e. $h<1$.
Both for the non-interacting case and interacting case, we find that the growth of entanglement with time obeys power-law type scaling i.e. $t^{\eta}$ (see Fig.~\ref{fig7}). This exponent $\eta$, seems to decrease with increasing $h$ (see Figure.~\ref{fig7}). Note that in the limit $h=0$ and $V=0$, this power-law growth of entanglement was reported in the context of Geometric quench in Ref.~\cite{gm2}. This feature is quite unique compared to the usual global quench, where the entanglement growth is always found to be linear in time in the delocalized phase~\cite{iyer.2013,kim.2013}, except for long-range systems ~\cite{modak.2020}.  However, the long-time saturation values of entanglement entropy followed by geometric quench for delocalized phase obey volume law (see the inset of Fig.~\ref{fig7}), which also has been observed for global quench. 
Now we focus on the localized phase. In the absence of interaction, the entanglement growth profile for geometric quench is very similar to the one is observed for the usual global quench, i.e. there is a short time growth followed by saturation, and the saturation values do not change with the system size, hence follows area law (see Fig.~\ref{fig8})~\cite{moore.2014}. However, the most striking results appear in the case of finite interaction. While in the usual global quench case, in the presence of interactions, the time evolution profile of the entanglement entropy is very different from the non-interacting case, the interacting case shows a logarithmic growth followed by a saturation, while saturation values of the entanglement entropy obey volume law. On the other hand, in the case of geometric quench, remarkably we find that the entanglement profile is very similar to the non-interacting results. The time evolution profile of the entanglement entropy does not possess a logarithmic growth, also the saturation value obeys area law (see Fig.~ \ref{fig8}).
\section{Conclusions}
In this work, our main goal was to understand the effect of Geometric quench on localized and delocalized phases. While there have been extensive studies on such systems, global quench has been used as a very important tool both for experimentally~\cite{mbl11,mbl12,mbl13} and theoretically to characterize these phases, the effect geometric quench has not been well explored so far. 
We use mainly two types of diagnostics 1) site-occupations profile and 2) entanglement entropy. While entanglement entropy is a very popular measure to detect localization-delocalization transition even when a system undergoes a global quench, site-occupation profile 
remains a useful tool for geometric quench. We have extensively studied the effect of the incommensurate potential strength and the interaction strength on the wavefront velocities. We found that in the delocalized phase the wavefront moves towards the boundary, but in the localized phase (even in the presence of interactions) the wavefront 
almost gets frozen i.e. almost no change can be observed even one waits a very long time.
On the other hand, the entanglement entropy shows quite a distinct feature compared to the global quench. In the delocalized phase, the entanglement growth is $t^{\eta}$ with $\eta <1$, in contrast to the linear growth is found in the case of global quench. However, the saturation values are observed to be obeying a volume i.e. same as the global quench. In the localized phase, the entanglement profile for the geometric quench can not be distinguished between Anderson localized phase and the MBL phase. In both cases, the saturation values obey Area law.

Given that the geometric quench can be experimentally realized in an ultra-cold setup~\cite{sudden_expt1,sudden_expt2}, our future plan will be to investigate similar protocols for long-range systems~\cite{modak.2020,modak.2020_1} and systems with single-particle mobility edges~\cite{modak.2018}. Also, it will be interesting to study non-Hermitian~\cite{modak.2021} systems in the shade of similar light. 

\section{Acknowledgements}
RM acknowledges DST-Inspire fellowship, by the Department of Science and Technology, Government of India, and 
SERB start-up grant (SRG/2021/002152) for their supports. The support and the resources provided by ‘PARAM Shivay Facility’ under the National Supercomputing
Mission, Government of India at the Indian Institute of Technology, Varanasi are gratefully acknowledged.

\end{document}